\documentclass[runningheads]{svmult}

\usepackage{makeidx}   % allows index generation
\usepackage{graphicx}  % standard LaTeX graphics tool
\usepackage{subeqnar}  % subnumbers individual equations
\usepackage{multicol}  % used for the two-column index
\usepackage{physprbb}  % modified textarea for proceedings,
\makeindex             % used for the subject index
\begin{document}
%
%own macro defs
\newcommand{\napikms}{km\,s$^{-1}$}

\title*{Search for double degenerate progenitors of supernovae type Ia 
with SPY\thanks{Based on data obtained at the Paranal Observatory of the
European Southern Observatory for programs 165.H-0588, 167.D-0407,
266.D-5658, 268.D-5739, 68.D-0483, 69.D-0534}
\thanks{Based on observations collected at the German-Spanish Astronomical
Center, operated by the Max-Planck-Institut f\"ur Astronomie Heidelberg
jointly with the Spanish National Comission for Astronomy}}
\toctitle{Search for double degenerate progenitors of supernovae type Ia 
with SPY}
% allows explicit linebreak for the table of content
%
%
\titlerunning{SPY}
% allows abbreviation of title, if the full title is too long
% to fit in the running head
%
\author{R.~Napiwotzki\inst{1}
\and N.~Christlieb\inst{2}
\and H.~Drechsel\inst{1}
\and H.-J.~Hagen\inst{2}
\and U.~Heber\inst{1}
\and D.~Homeier\inst{4}
\and C.~Karl\inst{1}
\and D.~Koester\inst{3}
\and B.~Leibundgut\inst{5}
\and T.R.~Marsh\inst{6}
\and S.~Moehler\inst{3}
\and G.~Nelemans\inst{7}
\and E.-M.~Pauli\inst{1}
\and D.~Reimers\inst{2}
\and A.~Renzini\inst{5}
\and L.~Yungelson\inst{8}}
% \author{Ivar Ekeland\inst{1}
\institute{Dr.~Remeis-Sternwarte, Astronom.\ Institut, Universit\"at 
        Erlangen-N\"urnberg, Sternwartstr.~7, 96049 Bamberg, Germany
\and Hamburger Sternwarte, Universit\"at Hamburg, Gojenbergsweg 112, 
  21029 Hamburg, Germany
\and Institut f\"ur Theoretische Physik und Astrophysik, 
  Universit\"at Kiel, 24098 Kiel, Germany
\and Department of Physics \& Astronomy,
        University of Georgia, Athens, GA\,30602-2451, USA
\and European Southern Observatory, Karl-Schwarzschild-Str.~2, 
  85748 Garching, Germany
\and University of Southampton, Department of Physics \& Astronomy, 
  Highfield, Southampton S017 1BJ, UK
\and Institute of Astronomy, Madingley Road, Cambridge CB3~0HA, UK
\and Institute of Astronomy of the Russian Academy of Sciences, 
  48 Pyatnitskaya Str., 109017 Moscow, Russia}
\authorrunning{Napiwotzki al.}
% if there are more than two authors,
% please abbreviate author list for running head
%
%

\maketitle              % typesets the title of the contribution

\begin{abstract}
We report on a large survey for double degenerate (DD)
binaries as potential progenitors of type Ia supernovae with the UVES
spectrograph at the ESO VLT (ESO {\bf S}N\,Ia {\bf P}rogenitor
surve{\bf Y} -- SPY). About 560 white dwarfs were checked for radial
velocity variations until now. Ninety new DDs have been discovered, including
short period systems with masses close to the Chandrasekhar mass.
\end{abstract}

\section{The project}
Supernovae of type Ia (SN\,Ia) play an outstanding role for our understanding
of galactic evolution and the determination of the extragalactic distance
scale.  However, the nature of their progenitors is still unknown (e.g.\ 
\cite{Liv00}).  There is general consensus that the event is due to the
thermonuclear explosion of a white dwarf when the Chandrasekhar mass
($1.4M_\odot$) is reached, but the nature of the progenitor system remains
unclear. Two main options exists: the merging of two WDs in the so called
double degenerate (DD) scenario \cite{IT84}, or mass transfer from a red
giant/subgiant in the so-called single degenerate (SD) scenario \cite{WI73}.
% with the system possibly appearing as a symbiotic binary \cite{MR92},
% supersoft X-ray source \cite{VBN92}, or recurrent nova \cite{HK01}. 

In the DD case, we know that most stars end up as white dwarf remnants and
that a major fraction of stars are in binary systems, hence DDs must be common
among WDs. What we do not know is whether there exist enough DDs able to merge
in less than one Hubble time and produce a SN\,Ia event. In the SD case, we
know that white dwarfs accreting from a non-degenerate companion do exist, but
we do not know whether such systems exist in sufficient number to account for
the observed SN\,Ia frequency, nor whether the WD grows enough to reach the
critical mass for ignition. There is also evidence from a class of
sub-luminous SN\,Ia that both routes might be significant \cite{How01}.

On the theoretical side three possible outcomes of the merger of a
super-Chandrasekhar mass DD are discussed in literature: a SN\,Ia explosion
leaving no remnant \cite{IT84},  \cite{Pie02}, an accretion induced
collapse producing a neutron star without a SN\,Ia explosion \cite{SN85}, or
the formation of a neutron star during a SN\,Ia event \cite{KPW01}.
On the observational side 
several systematic radial velocity (RV) searches for DDs have been
undertaken starting in the mid 1980's 
checking a total of $\approx 200$ white dwarfs RV for
variations (cf.\ \cite{Mar00} and references
therein),
but have failed to reveal any massive, short-period DD progenitor 
of SN\,Ia. This negative result casted some doubt on the DD scenario.
However, it is not unexpected, as theoretical simulations suggest that
 only a few percent of all DDs are potential SN\,Ia progenitors
\cite{ITY97}, \cite{NYP01}. 

In order to perform a definitive test of the DD scenario we have
embarked on a large spectroscopic survey  of $\ge 1000$ white dwarfs 
(ESO {\bf S}N \,Ia
{\bf P}rogenitor surve{\bf Y} -- SPY). 
SPY will overcome the main limitation of all efforts so far to detect
DDs that are plausible SN~Ia precursors: the samples of surveyed
objects were too small.  
Spectra were taken with the high-resolution 
UV-Visual Echelle Spectrograph (UVES) of
the UT2 telescope (Kueyen) of the ESO VLT in service mode. 
Our instrument setup 
%(Dichroic~1, central wavelengths 3900\,\AA\ and
%5640\,\AA, slit width 2.1$''$) 
provides nearly complete spectral coverage from 3200\,\AA\ to 
6650\,\AA\
%with only two $\approx$80\,\AA\ wide gaps at 4580\,\AA\ and 5640\,\AA\
with a resolution $R=18500$ (0.36\,\AA\ at H$\alpha$). 
Due to the nature of
the project, two spectra at different, ``random'' epochs separated 
by at least one day are observed.

ESO provides a data reduction pipeline for UVES, which
 formed the basis for our first selection
of DD candidates. A careful re-reduction of the spectra is in progress.
Differing from previous surveys we use a correlation procedure to 
determine RV shifts of the observed spectra 
(cf.\ \cite{NCD01}).
We routinely measure RVs with an accuracy of $\approx 2$\,\napikms\
or better, therefore running only a very small risk of missing a merger
precursor, which have orbital velocities of 150\,\napikms\ or higher.  

\section{Results.}
We have analyzed spectra of 558 white dwarfs and pre-white dwarfs
taken during the first two years
of the SPY project and detected 90 new DDs, 13 are double-lined systems (only 6
were known before). Results are summarized in Table~1.
%The great advantage of double-lined binaries is that they
%provide us with a well determined total mass.  
Our observations have already
increased the DD sample by a factor of six.  After completion, a final sample
of $\approx$200 DDs is expected.
SPY is the first RV survey which performs a systematic
investigation of both classes of white dwarfs: DAs {\it and} non-DAs. 
Previous surveys were restricted to DA white dwarfs, because the sharp NLTE
core of H$\alpha$ allows a very accurate RV determination. This feature is not
present in the non-DA (DB, DO) spectra, but the use of several helium-lines
enables us to reach a similar accuracy. Our results in Table~1 indicate that
the DD frequency among DA and non-DA white dwarfs are similar.
\begin{table}
\caption{Fraction of RV variable stars in the current SPY sample for 
different spectral classes. WD+dM denotes systems for which a
previously unknown cool companion is evident from the red spectra
(not included in the DA/non-DA entries).}
\label{t:DDs}

\begin{tabular}{l|r|r|r}
Spectral type   &total  &RV variable    &detection rate\\ \hline
All DDs    &558    &90     &16\%\\
non-DA (DB,DO,DZ)       &72     &10      &14\%    \\
WD+dM   &30     &14      &47\%   \\
\end{tabular}
\end{table}

Follow-up observations of this sample are mandatory to 
determine periods and white dwarf parameters
and find potential SN\,Ia progenitors among the candidates. 
Good statistics of a large DD sample will also set stringent
constraints on the evolution of close binaries, which will
dramatically improve our understanding of this phase of stellar
evolution.  
Starting in 2001 follow-up observations have been carried out with VLT 
and NTT of ESO as 
well as with the 3.5\,m telescope of the Calar Alto observatory/Spain
and the INT \cite{NCD01}, \cite{NEH01}, \cite{NKN02},
\cite{KNH02}). 
Our sample includes many short period binaries, several with masses closer to
the Chandrasekhar limit than any system known before.  
During our follow-up observations we have detected a very promising
potential SN\,Ia precursor candidate discussed below.

\begin{figure}
\includegraphics[width=.95\textwidth]{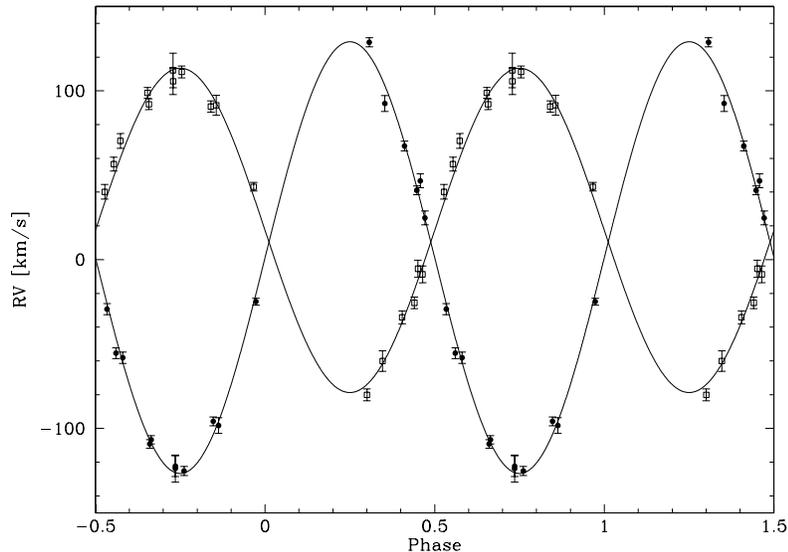}
\caption{Measured RVs as a function of orbital phase and
fitted sine curves for HE\,1414-0848. Filled circles/open rectangles 
 indicate the less/more massive components A and B. 
Note the difference of the
``systemic velocities'' $\gamma_0$ between both components caused by
gravitational redshift.}
\label{f:HE1414rv}
\end{figure}

Our follow-up observations concentrated on candidates with high RV variations,
indicating short periods. Double-lined systems (with both white dwarfs visible
in the spectrum) are of special interest, because these binaries allow the
determination of individual masses for both components.

Exemplary for other double-lined systems we discuss here the DA+DA system
HE\,1414$-$0848 \cite{NKN02}.  
The orbital period of $P = 12^{\mathrm{h}} 25^{\mathrm{m}}
44^{\mathrm{s}}$ and semi-am\-pli\-tudes of 
127\,km\,s$^{-1}$ and 96\,km\,s$^{-1}$
are derived for the individual components.  RV curves for both
components are displayed in Fig.~\ref{f:HE1414rv}.  The ratio of velocity
amplitudes is directly related to the mass ratio of both components.
Additional information comes from the mass dependent gravitational redshift.
%$z =\Delta \lambda/\lambda= GM R^{-1}c^{-2}$, which for 
% a given mass-radius relation
%can be computed as a function of white dwarf mass. 
The difference in gravitational redshift corresponds to the apparent
difference of ``systemic velocities'' of both components, as derived from the
RV curves (Fig.~1).  Only one set of individual white dwarf masses fulfills
the constraints given by both the amplitude ratio and redshift difference (for
a given mass-radius relation).  We estimate the masses of the individual
components with this method to be $0.55M_\odot$ and $0.71M_\odot$ for A and B,
respectively.  This translates into $\log g = 7.92$ and 8.16, respectively.
 
\begin{figure}
\includegraphics[width=0.95\textwidth]{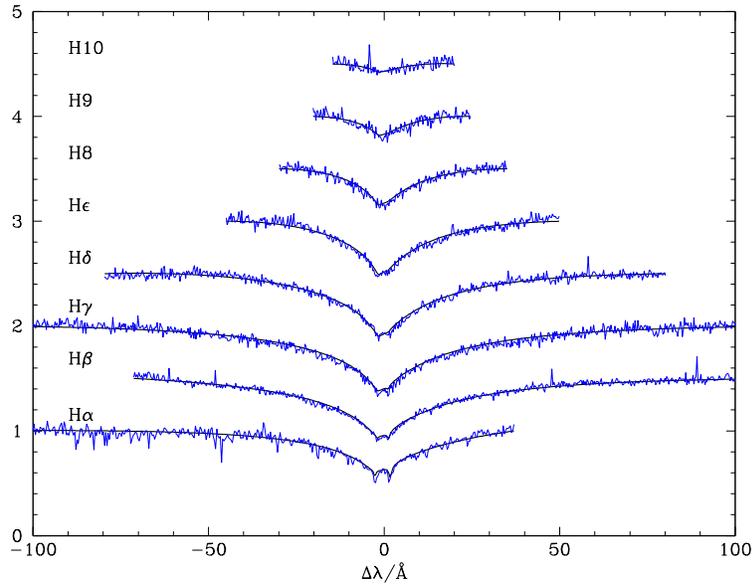}
\caption{Model atmosphere fit of the Balmer series of HE\,1414$-$0848 with
  FITSB2. This is only a sample fit. All available spectra, covering different
  orbital phases, were used, which allows an unambiguous parameter
  determination.}
\label{f:HE1414fit}
\end{figure}

Another estimate of the white dwarf parameters is available from a model
atmosphere analysis of the combined spectrum. We have developed a new tool
({\sc FITSB2}), which performs a spectral analysis of both components of a
double-lined system. The fit is performed on all available spectra, covering
different orbital phases simultaneously.  We fitted temperatures and gravities
of both components of HE\,1414$-$0848 (the mass ratio fixed at the accurate
value derived from the RV curve).  The results are $T_{\mathrm{eff}}$/$\log g$
= 8380\,K/7.83 and 10900\,K/8.14 for A and B (Fig.~2), which are in good
agreement with the $\log g$ values predicted from the analysis of the RV
curve.  The total mass of the HE\,1414$-$0848 system is $1.26M_\odot$, only
10\% below the Chandrasekhar limit.  The system will merge due to loss of
angular momentum via gravitational wave radiation after two Hubble times.

\begin{table}
\caption{New double-lined DDs from SPY (data taken from \cite{NKN02} and
  \cite{KNH02}). The table details the mass sum ($M_1+M_2$) of the systems, 
  the orbital periods
  ($P$), and
  the time ($T_{\mathrm{merge}}$) until merging due to gravitational wave 
  radiation.}
\label{t:SB2}

\begin{tabular}{l|l|r|r}
\multicolumn{1}{c|}{system}     &$M_1+M_2$      &\multicolumn{1}{c|}{$P$}
        &$T_{\mathrm{merge}}$ \\
        &\multicolumn{1}{c|}{[$M_\odot$]} &\multicolumn{1}{c|}{h} &Gyrs \\ 
        \hline
HE\,1414$-$0848   &1.26   &12.4     &25\\
HE\,2209$-$1444   &1.26   &6.6      &4\\
WD\,1349+144      &0.88   &53.0    &2000\\
\end{tabular}

\end{table}

Our follow-up observations yielded parameters for several DDs. We have
completed the analysis for three double-lined systems summarized in
Table~\ref{t:SB2}. The mass sum of the HE\,2209$-$1444 system is as high as
that of HE\,1414$-$0848, but the HE\,2209$-$1444 system 
is closer and will merge
within 4\,Gyrs. On the other hand the WD\,1349+144 system 
consists of two low mass white dwarfs and has a rather long period and will
merge only after more than 100 Hubble times.

Our sample includes also another
 short period ($P=7^{\mathrm{h}}12^{\mathrm{min}}$)
system, which will merge after 4\,Gyrs and has probably a system mass above
the Chandrasekhar limit. Thus this system is a very promising potential SN\,Ia
precursor candidate. However, the H$\alpha$ line core of the secondary in this
system is broad and shallow compared to typical DA line cores, which makes the
determination of a precise RV curve and therefore of the masses difficult.
Some additional data are necessary to verify our RV curve solution.  Results
will be reported elsewhere.

\section{Conclusions}

SPY has now finished a major fraction of the survey. Our analysis of the data 
from the first 24 months has already quadruplicated the number of
white dwarfs checked for RV variability (from 200 to 760) and increased the
number of known DDs from 18 to 108 compared to the results of the last 20
years. Our sample includes many short period systems (Table~1; \cite{NEH01},
\cite{NKN02}, \cite{KNH02}), several with masses closer to the Chandrasekhar
limit than any system known before, greatly improving the statistics of
DDs. We expect this survey to produce a final sample of $\approx$200 DDs. 

This will also provide a census of the final binary configurations, hence an
important test for the theory of close binary star evolution after mass and
angular momentum losses through winds and common envelope phases, which are
very difficult to model. An empirical calibration provides the most promising
approach. A large sample of binary white dwarfs covering a wide range in
parameter space is the most important ingredient for this task. We have
started a project to exploit the information provided in the SPY sample
of DDs.
% Predictions from population synthesis calculations \cite{ITY97}, \cite{NYP01} 
% are compered with the properties of the observed sample and will be used to 
% improve the models. 

Our ongoing follow-up observations already revealed the existence of
three short period systems with masses close to the
Chandrasekhar limit, which will merge within 4\,Gyrs to two Hubble times. 
Even if it will finally turn out that the mass of our most
promising SN\,Ia progenitor candidate system is slightly below the
Chandrasekher limit, our results already  allow
 a qualitative evaluation of the DD channel. 
Since the formation of a system slightly below the Chandrasekhar limit is not
very different from the formation of a system above this limit, the presence
of these three systems alone provides evidence that potential DD progenitors of
SN\,Ia do exist.

% \paragraph{Spin-off results.} SPY produces an immense, unique
% sample of very high resolution white dwarf spectra. It will allow us for
% the first time to tackle many longstanding questions on a firm
% statistical basis.  Among those are the mass distribution of white dwarfs
% (\cite{KNC01}), 
% kinematical properties of the white dwarf population (Pauli, these
% proceedings), surface compositions,
% luminosity function, rotational velocities, and detection of weak
% magnetic fields. 
% A more detailed description of ongoing spin-off activity
% is given in \cite{NCD01}. 
% Members of the community interested in spin-off opportunities are invited
% to participate in the exploitation of the SPY sample.

%INDEX%%%%%%%%%%%%%%%%%%%%%%%%%%%%%%%%%%%%%%%%%%%%%%%%%%%%%%%%%%%%%%%
% Please check with the editor of your book whether he plans to
% include a "mutual" subject index - if so, please code your entries
% in the standard syntax. For your own purposes you may print your
% "personal" index by using the following commands:
%
%\clearpage
%\addcontentsline{toc}{section}{Index}
%\flushbottom
%\printindex
%%%%%%%%%%%%%%%%%%%%%%%%%%%%%%%%%%%%%%%%%%%%%%%%%%%%%%%%%%%%%%%%%%%%%

\end{document}